\documentclass[conference]{IEEEtran}
\IEEEoverridecommandlockouts                             
\usepackage{cite}
\usepackage{amsmath,amssymb,amsfonts}
\usepackage{graphicx}
\usepackage{subcaption}
\usepackage{float}           % this is required for better table positioning
\newtheorem{remark}{Remark}  % this is required for numbered Remarks
\title{\LARGE\bf	Sign Gradient Descent Algorithms for Kinetostatic Protein Folding}
\author{A. Mohammadi$^{1\ast}$, and M. Al Janaideh$^{2}$
	\thanks{$^{1}$A. Mohammadi is with the Department of Electrical \& Computer Engineering, University of Michigan, Dearborn, 4901 Evergreen Road, MI 48128, USA. $^{2}$ M. Al Janaideh is with the Department of Mechanical Engineering, University of Guelph, Guelph, ON  N1G 2W1, Canada. This work is supported by the National Science Foundation (NSF) through the award number CMMI-2153744.} 
	\thanks{$^{\ast}$Corresponding Author: A. Mohammadi, Email: {\tt\small amohmmad@umich.edu}}%
}
\begin{document}
	\maketitle
	\begin{abstract}
		%% Text of abstract
		This paper proposes a sign gradient descent (SGD) algorithm for predicting the three-dimensional folded protein molecule structures under the kinetostatic compliance method (KCM).  In the KCM  framework, which can be used to simulate the range of motion of peptide-based nanorobots/nanomachines, protein molecules are modeled as a large number of rigid nano-linkages that form a kinematic mechanism under motion constraints imposed by chemical bonds while folding under the kinetostatic effect of nonlinear interatomic force fields.  In a departure from the conventional successive kinetostatic fold compliance framework, the proposed SGD-based iterative algorithm in this paper results in convergence to the local  minima of the free energy of protein molecules corresponding to their final folded conformations in a faster and more robust manner. KCM-based folding dynamics simulations of the backbone chains of protein molecules demonstrate the effectiveness of the proposed algorithm.  
	\end{abstract}

\section{Introduction}
Numerical simulations that aim at providing a prediction of the  three-dimensional structure of folded protein conformations and computing the transitions through which these molecules fold/unfold play an integral role in designing protein-based nanomachines/nanorobots. Indeed, such numerical simulations can estimate the range of motion of these peptide-based mechanisms. For instance, the design of parallel nanorobots, which  consist of graphite platforms interconnected together via serially linked protein-based bio-springs, requires numerical simulations for finding the motion pattern of the linear protein actuators within the nano-mechanism (see, e.g.,~\cite{hamdi2005molecular,hamdi2009multiscale}).

One class of algorithms for predicting the final folded structures of protein molecules  is afforded by the so-called knowledge-based approaches. Rooted in pattern recognition and machine learning, these algorithms predict three-dimensional structures of folded protein conformations by considering the linear amino acid sequence of a given protein molecule and utilizing  massive datasets of already available folds~\cite{moult2014critical}. The family of knowledge-based solutions, to which the Google AlphaFold~\cite{alquraishi2020watershed} belongs, cannot capture the protein-nucleic acid interactions and address the computation of folding pathways, namely, the transient conformations~\cite{fersht2021alphafold} through which the protein molecule attains its folded conformation. Indeed, AI-based methods have been able to find the most likely folded  conformations without considering the stability and kinetics of the  folding process. 

\begin{figure}
	\centering
	\includegraphics[width=0.42\textwidth]{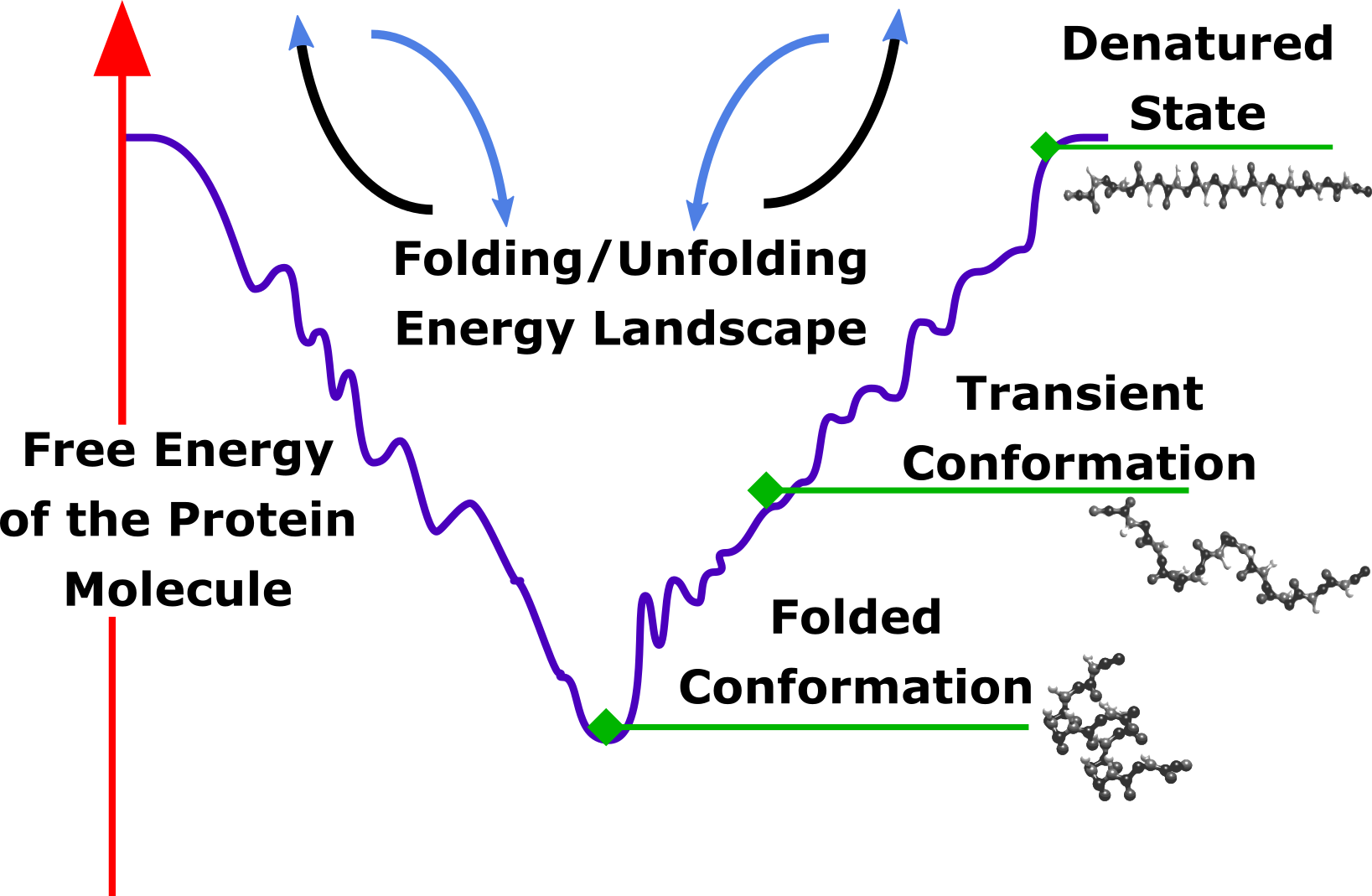}
	\caption{\small Protein folding/unfolding against the free energy landscape of the molecule. In the KCM framework for protein folding~\cite{kazerounian2005protofold,tavousi2015protofold}, the molecule dihedral angles vary under the nonlinear effect of interatomic  force fields resulting in protein conformational changes until convergence to a minimum energy state.}
	\vspace{-0.0ex}
	\label{fig:protBasic2}
\end{figure}

The counterpart to knowledge-based approaches, namely, the family of physics-based methods, relies on using  physical first-principles to numerically compute the folding pathways and predict the final three-dimensional folded  structure of protein molecules~\cite{heo2019driven,oldziej2005physics,fersht2002protein}. To increase the accuracy of folding pathway computations, numerous optimization techniques such as optimal control-based~\cite{arkun2012combining,arkun2011protein} and homotopy-based~\cite{dunlavy2005hope} algorithms can also be  augmented with these numerical methods.  While physics-based approaches provide reliable information about the transient conformations during the folding process, they are computationally burdensome.

The promising framework of kinetostatic compliance method (KCM), pioneered by Kazerounian, Ilie{\c{s}}, and collaborators,  models protein molecules as a large number of rigid nano-linkages that form a kinematic mechanism under motion constraints imposed by chemical bonds~\cite{kazerounian2005protofold,tavousi2015protofold,behandish2013gpu,madden2009residue}. In this framework, which addresses the high computational load of all-atom molecular dynamics approaches, the dihedral angles, which determine the molecule three-dimensional structure, change under the nonlinear effect of interatomic  force fields resulting in protein conformational changes until convergence to a minimum energy state. A schematic of protein folding/unfolding against the free energy landscape of the molecule is depicted in Figure~\ref{fig:protBasic2}.

Since its advent, the KCM framework has been successfully utilized for investigation of the role of hydrogen bond formation in protein kinematic mobility problem~\cite{shahbazi2010} and design of peptide-based nanomachines~\cite{mundrane2022exploring,chorsi2020one,Tavousi2016}.  For instance, Mundrane \emph{et al.}~\cite{mundrane2022exploring} have utilized the KCM framework for simulating the range of motion for closed-loop cyclic 7-R peptides that are subject to external electric field perturbations. Moreover, it has been demonstrated that  entropy-loss constraints during folding can be encoded in the KCM framework by using a proper nonlinear optimization-based control  scheme~\cite{mohammadi2021quadratic}. Moreover, the KCM framework can be used for systematic investigation of the reverse process of folding, namely, protein unfolding~\cite{mohammadi2022chetaev}. 

Despite the KCM  computational advantages for numerical simulation of protein folding dynamics and its utilization in design of peptide-based nanorobots/nanomachines, this framework has exclusively relied on the so-called \emph{successive kinetostatic fold compliance} (see, e.g.,~\cite{kazerounian2005protofold,tavousi2015protofold}), where the iterative conformational changes of the protein molecule are computed by taking steps along a special direction determined by heuristics. Furthermore, the convergence properties of the successive kinetostatic fold compliance has not been investigated in the literature due to its heuristic nature. 

In this paper, we examine the heuristics behind the conventional successive kinetostatic fold compliance for protein folding dynamics and arrive at a sign gradient descent (SGD) iterative algorithm as an alternative to the conventional approach. SGD algorithms, which were originally proposed in the context of training artificial neural networks (see, e.g.,~\cite{pascanu2013difficulty}), are a class of first-order methods that merely involve the sign of the gradient of the objective function (the free energy of the protein molecule in our case) while enjoying numerical stability and robust convergence properties~\cite{kolter2009policy,moulay2019properties}. In robotics applications, SGD algorithms have been utilized in settings such as autonomous environmental monitoring where the sign of the change in gradient (not the magnitude of the change) plays a crucial role in planning the motion of the robot (see, e.g.,~\cite{dhariwal2004bacterium}). 

\noindent\textbf{Contributions of the paper.} This paper contributes to the KCM-based protein folding framework by developing a family of SGD algorithms for  numerical simulation of protein folding dynamics. This contribution is a departure from the established literature (see, e.g.,~\cite{kazerounian2005protofold,tavousi2015protofold}) of the KCM folding framework where numerical simulations of the folding dynamics have exclusively relied on the heuristic successive kinetostatic fold compliance scheme. Moreover, by relying on the rich literature of SGD optimization (see, e.g.,~\cite{moulay2019properties}), this paper provides formal conditions under which the proposed numerical SGD-based iterative algorithm for kinetostatic folding converges to folded protein  conformations. Finally,  the proposed SGD-based iterative algorithm in this paper results in convergence to the local  minima of the free energy of protein molecules corresponding to their final folded conformations in a faster and more robust manner.  

% organization
The rest of the paper organization is as follows. First, in Section~\ref{sec:dynmodel}, we provide an overview of the kinematics of protein molecules and the KCM framework for modeling the protein folding process. Thereafter, in Section~\ref{sec:importance}, we present the conventional KCM-based iteration and our SGD-based alternative to it.  The numerical simulation results are presented in Section~\ref{sec:sims}. Finally, the paper is  concluded with future research directions and final remarks in Section~\ref{sec:conclusion}. 
\noindent{\textbf{Notation.}} We denote the set of all non-negative real and non-negative integer numbers by $\mathbb{R}_{+}$ and $\mathbb{Z}_{0+}$, respectively.  Given a positive integer $M$, a vector $\mathbf{x}=[x_1,\cdots,x_M]^\top$ in $\mathbb{R}^M$, and a real constant $p\geq 1$, we denote the $p$-norm of the vector by $|\mathbf{x}|_p$. Furthermore, we let $|\mathbf{x}|_\infty=\max\limits_{i} |x_i|$. We denote the sign function by $\text{sgn}(\cdot)$, which is defined according to $\text{sgn}(a)=1$ if $a>0$, $\text{sgn}(a)=0$ if $a=0$, and $\text{sgn}(a)=-1$ if $a<0$. Given a vector-valued function $\mathbf{f}(\mathbf{x})=[f_1(\mathbf{x}),\cdots,f_M(\mathbf{x})]^\top$ for some positive integer $M$,  we denote  $\text{sgn}(\mathbf{f}(\mathbf{x}))=\big[ \text{sgn}(f_1(\mathbf{x})),\cdots, \text{sgn}(f_M(\mathbf{x})) \big]^\top$. 

\section{Kinetostatic Compliance-Based Protein Folding}
\label{sec:dynmodel} 
% Review of KCM approach to protein folding; comparison with targeted molecular dynamics and turnign the iterations to 
% ODEs (setting the ground work for numerical integrators)
%
In this section, we present an overview of the KCM framework for modeling the \emph{in vacuo} folding dynamics of protein molecules. 

\subsection{Nano-linkage-based kinematic model of protein molecules}
\label{subsec:nanoLinkage}
Protein molecules are long molecular chains that consist of peptide planes with peptide chemical bonds joining them together. For brevity, we limit our presentation to the protein main chain. Indeed, the essential folding dynamics can be effectively explained by considering the motion of the protein backbone chain (see, e.g.,~\cite{amadei1993essential}). 

%The protein side chains, nonetheless, are required to explain phenomenon such as hydrophobic core collapse~\cite{shao2010effects}. The readers are referred to~\cite{tavousi2015protofold,kazerounian2005protofold,kazerounian2005nano} for the details on how to model the effect of protein side chains.

\begin{figure*}[t]
	\centering
	\includegraphics[width=0.65\textwidth]{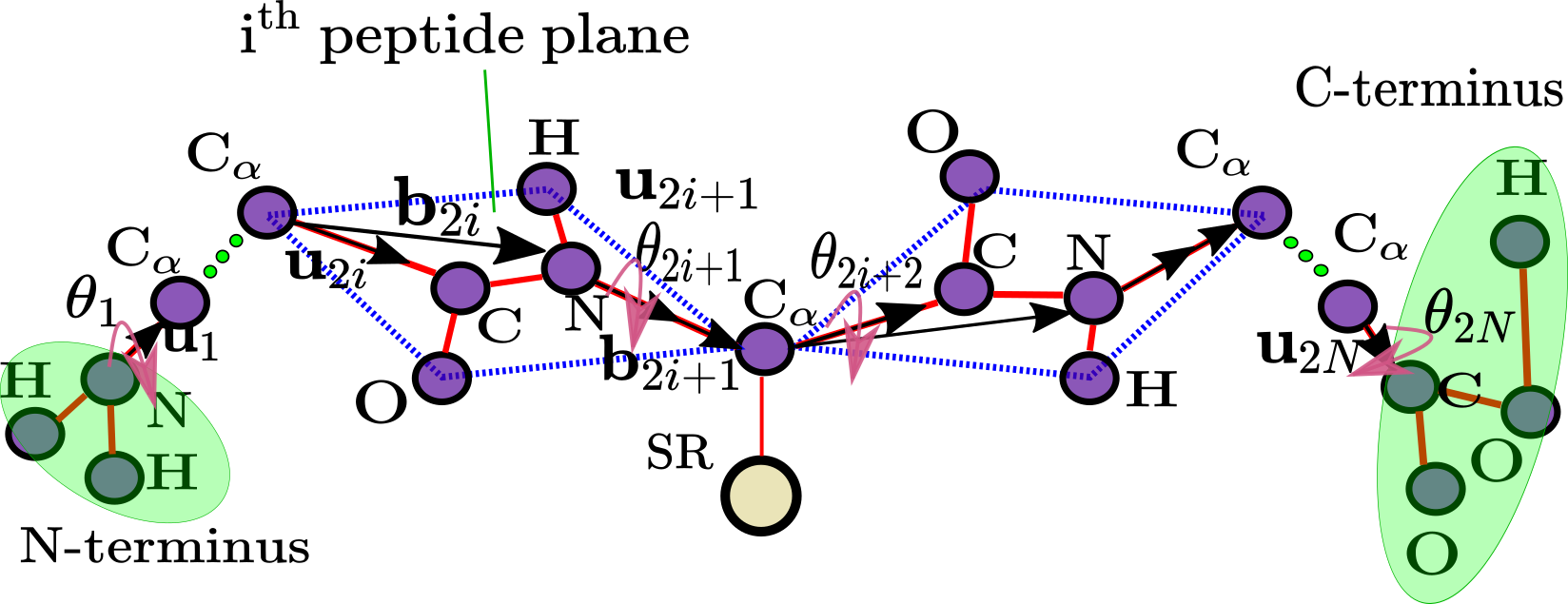}
	\caption{\small The protein molecule kinematic mechanism consisting of peptide planes  and $\text{C}_\alpha$ atom hinges. There also exists a hydrogen atom, which is not depicted in the figure,  connected to each $\text{C}_\alpha$ atom via a covalent bond.}
	\vspace{-0.5ex}
	\label{fig:protBasic}
\end{figure*}

As demonstrated in Figure~\ref{fig:protBasic}, each peptide plane, which consists of six coplanar atoms, can be considered as a linkage in the protein kinematic  mechanism~\cite{kazerounian2005nano}. Central carbon atoms, which are denoted by $\text{C}_\alpha$ and commonly known as the alpha-Carbon atoms, act as hinges connecting peptide planes together.  The peptide plane atoms are bonded together via covalent bonds (i.e., the red line segments in Figure~\ref{fig:protBasic}). 
\begin{remark}
	\label{rem:coplanar}
	The assumption of coplanarity of the atoms $\text{C}_\alpha$, $\text{CO}$, $\text{NH}$, and $\text{C}_\alpha$, which form each of the peptide planes (see Figure~\ref{fig:protBasic}), is based on the results from high resolution X-ray crystallographic experiments (see, e.g.,~\cite{finkelstein2016protein}). This coplanarity assumption has been the basis of various robotics-inspired approaches in  the literature that model protein molecules as robotic mechanisms with hyper degrees-of-freedom (see, e.g.,~\cite{tavousi2015protofold,kazerounian2005nano,mohammadi2021quadratic,mohammadi2022chetaev}). 
\end{remark}

Each alpha-carbon atom is bonded to four other chemical components including the three atoms $\text{C}$, $\text{N}$, and $\text{H}$, and a variable side chain shown with SR. The first alpha-Carbon of the protein chain structure is bonded to N-terminus, which is an amino group, as well as one  peptide plane.  Similarly, the last $\text{C}_\alpha$ atom is chemically bonded to the C-terminus, which is a carboxyl group, as well as one other peptide plane.  

The backbone conformation of the protein molecule kinematic structure consisting of the subchain $-\text{N}-\text{C}_\alpha-\text{C}-$, is described by a collection of 
bond lengths and a set of pairs of dihedral angles, namely, the angles representing the rotations around the covalent bonds 
$\text{C}_\alpha-\text{C}$ and $\text{N}-\text{C}_\alpha$ (see Figure~\ref{fig:protBasic}). Accordingly, 
\begin{equation}
\label{eq:thetaConfig}
\pmb{\theta}:= \big[\theta_1,\cdots,\theta_{2N}]^\top, 
\end{equation}
is the configuration vector of the kinematic structure of a given protein backbone chain with  $N-1$ peptide planes. 

\begin{remark}
	In the biochemistry literature, `conformation' is the standard word for describing the geometry of the protein molecule kinematic structure. In the robotics literature, on the other hand, the terminology `configuration' is frequently used to describe the kinematic structures of  robots. In this paper, unless otherwise stated, we use the two words `conformation' and `configuration' interchangeably.
\end{remark}

%As indicated by~\eqref{eq:thetaConfig}, the vector $\pmb{\theta}$ of dihedral angles  belongs %to the $2N$-dimensional configuration space $\mathcal{Q}:=\mathcal{S}^1\times \cdots \times %\mathcal{S}^1$, where $\mathcal{S}^1$ is the unit circle. 

Each of the dihedral angles in the conformation vector $\pmb{\theta}$ in~\eqref{eq:thetaConfig} correspond to one degree-of-freedom (DOF)   
of the protein molecule kinematic chain. Associated with each DOF, one may consider a unit vector denoted by 
$\mathbf{u}_j$, $1\leq j \leq 2N$. Each of these vectors are aligned with the rotation axis about which the kinematic chain can rotate. 
Therefore, as demonstrated in Figure~\ref{fig:protBasic}, the vectors $\mathbf{u}_{2i}$ and $\mathbf{u}_{2i+1}$ represent the unit vectors along the $\text{C}_\alpha-\text{C}$ bond and $\text{N}-\text{C}_\alpha$ bond  located within the $i$-th peptide plane, respectively. Finally, $\mathbf{u}_1$ and $\mathbf{u}_{2N}$ are the unit vectors of the $\text{N}$- (the amino group) and $\text{C}$-termini (the carboxyl group), respectively. 

An additional collection of vectors, which are called the \textbf{body vectors}, are required  to completely determine the spatial orientation of the rigid peptide nano-linkages in protein molecules. The body vectors, which  are denoted by $\mathbf{b}_{j}$, $1\leq j \leq 2N$, complete the  description of the relative position of the coplanar atoms that are located within each of the peptide planes. Specifically, the  relative position of any two atoms is given by a linear sum of the form $k_{1m} \mathbf{b}_{2i} + k_{2m} \mathbf{b}_{2i+1}$, in which the coefficients $k_{1m}$ and $k_{2m}$, $1\leq m \leq 4$, are constant and the same across all peptide linkages (see, e.g.,~\cite{Tavousi2016} for further details).  

The body vectors $\mathbf{b}_{j}$ along with the unit vectors $\mathbf{u}_j$ can be utilized to provide a complete description of the protein molecule conformation as a function of the dihedral angle vector $\pmb{\theta}$ consisting of the peptide dihedral angles. Indeed, after one designates the zero position configuration with $\pmb{\theta}=\mathbf{0}$, the  matrix transformations 
\begin{equation}
\mathbf{u}_j(\pmb{\theta}) = \Xi(\pmb{\theta},\mathbf{u}_j^0) \mathbf{u}_j^0  ,\, 
\mathbf{b}_j(\pmb{\theta}) = \Xi(\pmb{\theta},\mathbf{u}_j^0) \mathbf{b}_j^0,
\label{eq:kineProtu}
\end{equation}
determines the kinematic structure of the protein molecule using the dihedral angle conformation vector $\pmb{\theta}$. In~\eqref{eq:kineProtu}, the transformation matrix $\Xi(\pmb{\theta},\mathbf{u}_j^0)$ is defined according to 
\begin{equation}
\Xi(\pmb{\theta},\mathbf{u}_j^0):=\prod_{r=1}^{j} R(\theta_j,\mathbf{u}_j^0). 
\label{eq:kineProtu0}
\end{equation}
Furthermore, in~\eqref{eq:kineProtu0}, the rotation about the vector $\mathbf{u}_j^0$ with angle $\theta_j$ is given by the rotation matrix $R(\theta_j,\mathbf{u}_j^0)$. After determining the body  vectors $\mathbf{b}_j(\pmb{\theta})$ from~\eqref{eq:kineProtu} and fixing the N-terminus atom at the  origin, the Cartesian coordinates of the $k$\textsuperscript{th}-peptide plane atoms are given by 
\begin{equation}
\mathbf{r}_i(\pmb{\theta}) = \sum_{j=1}^{i}  \mathbf{b}_j(\pmb{\theta}), \;\; 1 \leq i \leq 2N - 1,  
\label{eq:kineProtu2}
\end{equation}
where the integers $i=2k-1$ and $i=2k$ represent the nitrogen atoms and the alpha-Carbon atoms, respectively. % of the $k$\textsuperscript{th}-plane
\subsection{KCM-based dynamics of protein folding}
The KCM approach for modeling the protein folding process pioneered by Kazerounian and collaborators is based on the well-established fact that the essential folding dynamics can be explained by neglecting the inertial forces (see, e.g.,~\cite{adolf1991brownian,kazerounian2005protofold,arkun2010prediction,arkun2012combining}).  Instead, the dihedral angles vary kinetostatically under the effect of the interatomic  force fields. Consequently, the dihedral angles  at each conformation of the protein molecule change in proportion to the effective torques acting  on the peptide chain. 

Considering a peptide chain with $N_a$ atoms and $N-1$ peptide planes, where the dihedral angle vector is given by $\pmb{\theta}$ in~\eqref{eq:thetaConfig},  and denoting the Cartesian position of any two atoms $a_i$, $a_j$ in the protein chain by $r_i(\pmb{\theta})$, $r_j(\pmb{\theta})$, their distance can be computed from $d_{ij}(\pmb{\theta}):=  \lvert r_i(\pmb{\theta})-r_j(\pmb{\theta})\rvert$. The parameters associated with respective electrostatic charges of the atoms in the protein molecule, the van der Waals radii of these atoms,  the van der Waals distance between any two atoms, their dielectric constant, the depth of potential well of any pair of atoms, and the weight factors for the electrostatic and van der Waals forces between any pair of two atoms can be found from~\cite{Tavousi2016} and the references therein.  Under these considerations, the aggregated free energy of the protein molecule can be written as 
\begin{equation}\label{eq:freePot}
\mathcal{G}(\pmb{\theta}) := \mathcal{G}^{\text{elec}}(\pmb{\theta}) + \mathcal{G}^{\text{vdw}}(\pmb{\theta}),
\end{equation}
where $\mathcal{G}^{\text{elec}}(\pmb{\theta})$ and $\mathcal{G}^{\text{vdw}}(\pmb{\theta})$ 
are the protein molecule electrostatic potential energy and the van der Waals interatomic potential energy, respectively (see, e.g.,~\cite{tavousi2015protofold} for the detailed expressions). The resultant forces of Coulombic and van der Waals nature  exerted on each atom $a_i$, $1\leq i \leq N_\text{a}$, can be computed from $F_{i}^{\text{elec}}(\pmb{\theta})=-\nabla_{\mathbf{r}_i} \mathcal{G}^{\text{elec}}$ and $F_{i}^{\text{vdw}}(\pmb{\theta})=-\nabla_{\mathbf{r}_i}\mathcal{G}^{\text{vdw}}$, respectively. 
% \begin{equation}\label{eq:vdw}
% \mathcal{G}^{\text{vdw}}(\pmb{\theta}) =  \sum_{i=1}^{N_\text{a}} \sum_{j\neq i} w_{ij}^{\text{vdw}}\epsilon_{ij} \bigg[\big(\frac{D_{ij}}{d_{ij}(\pmb{\theta})}\big)^{12}-2\big(\frac{D_{ij}}{d_{ij}(\pmb{\theta})}\big)^{6}\bigg],
% \end{equation}
%
% \noindent is  in the protein molecule.  

According to the KCM-based modeling framework~\cite{tavousi2015protofold}, it is required to compute the resultant forces and torques acting on the peptide planes in the protein molecule. Subsequently, the computed forces and torques are appended in the $6N$-dimensional vector $\mathcal{F}(\pmb{\theta})$, which is the generalized force vector directing the process of protein folding. In the next step, the generalized force vector $\mathcal{F}(\pmb{\theta})$ needs to be mapped to an equivalent torque vector, which is responsible for varying the dihedral angle vector of the protein molecule.   Specifically, the vector $\pmb{\tau}(\pmb{\theta})\in \mathbb{R}^{2N}$, which is along the gradient of the aggregated free energy $\mathcal{G}(\pmb{\theta})$, is given by 
\begin{equation}\label{eq:tau_KCM}
\pmb{\tau}(\pmb{\theta}) = \mathcal{J}^\top(\pmb{\theta}) \mathcal{F}(\pmb{\theta}), 
\end{equation}
\noindent where the matrix  $\mathcal{J}(\pmb{\theta})\in \mathbb{R}^{6N\times 2N}$ 
is the molecule chain Jacobian at conformation $\pmb{\theta}$ (see~\cite{tavousi2015protofold} for further details). 

At each folded protein molecule conformation $\pmb{\theta}^\ast$, which corresponds to a local minimum of the aggregated free energy $\mathcal{G}(\pmb{\theta})$, the torque vector vanishes, namely, $\pmb{\tau}(\pmb{\theta}^\ast)=\mathbf{0}$. As described in the next section, although the torque vector $\pmb{\tau}(\pmb{\theta})$ is along the steepest-descent direction of the free energy gradient in the conformation landscape, Kazerounian and collaborators~\cite{kazerounian2005nano,tavousi2015protofold} have noticed that using a normalized version of the torque vector for iterative update of the protein molecule conformations would have an improved performance in terms of stability and convergence rate.

\section{The Conventional KCM-based Iteration and its SGD-based Alternative}
\label{sec:importance}

In this section we first present the conventional KCM-based iteration for protein folding dynamics. Next, by closely examining the heuristics behind this numerical scheme, we propose an  SGD-based successive kinetostatic fold compliance alternative and present its convergence properties. 

Given an unfolded protein molecule conformation $\pmb{\theta}_0$, the conventional successive kinetostatic fold compliance, which relates the joint torques to the changes in the dihedral angles, is given by the numerical iteration (see, e.g.,~\cite{kazerounian2005protofold,tavousi2015protofold})
\begin{equation}
\pmb{\theta}_{k+1} = \pmb{\theta}_k + \kappa_0\, \frac{\pmb{\tau}(\pmb{\theta}_k)}{| \pmb{\tau}(\pmb{\theta}_k) |_\infty}, \; k\in \mathbb{Z}_{0+}, 
\label{eq:KCMconv}
\end{equation}
where the normalized torque vector $\tfrac{\pmb{\tau}(\pmb{\theta}_k)}{| \pmb{\tau}(\pmb{\theta}_k) |_\infty}$  in~\eqref{eq:KCMconv} is responsible for varying the dihedral angle vector  at each conformation $\pmb{\theta}_k$. Moreover, the positive constant $\kappa_0$ is chosen small enough to avoid large variations in the dihedral angles and is tuned in a heuristic manner. The iterative steps in~\eqref{eq:KCMconv} are repeated until   the aggregated free energy $\mathcal{G}(\pmb{\theta})$ of the molecule  converges to a close vicinity of a free-energy-landscape local minimum where the norm of the torque vector, namely,  $|\pmb{\tau}(\pmb{\theta}_k)|_2$, becomes less than a desired tolerance $\tau_\text{tol}$ (i.e., the convergence criterion is met if $|\pmb{\tau}(\pmb{\theta}_k)|_2 < \tau_\text{tol}$).

Despite the fact that the torque vector $\pmb{\tau}(\pmb{\theta}_k)$ in the conventional KCM-based iteration given by~\eqref{eq:KCMconv} is along the steepest-descent direction of the free energy gradient in the conformation landscape, Kazerounian and collaborators~\cite{kazerounian2005nano,tavousi2015protofold} have noticed that using the normalized torque vector $\tfrac{\pmb{\tau}(\pmb{\theta}_k)}{| \pmb{\tau}(\pmb{\theta}_k) |_\infty}:=\big[\tfrac{\tau_1(\pmb{\theta}_k)}{| \pmb{\tau}(\pmb{\theta}_k) |_\infty},\cdots,\tfrac{\tau_{2N}(\pmb{\theta}_k)}{| \pmb{\tau}(\pmb{\theta}_k) |_\infty}\big]^\top$ for iterative update of the protein molecule conformations would outperform using $\pmb{\tau}(\pmb{\theta}_k)$ in terms of stability and convergence rate. Indeed, normalizing by the maximum joint torque $| \pmb{\tau}(\pmb{\theta}_k) |_\infty = \max\limits_{i} | \tau_i(\pmb{\theta}_k) |$ throughout the entire chain, results in normalizing the torques according to $\frac{\tau_i(\pmb{\theta}_k)}{| \pmb{\tau}(\pmb{\theta}_k) |_\infty} \in [-1,1]$. 

\emph{The aforementioned analysis leads to the important insight that the magnitude of the torque vector $\pmb{\tau}(\pmb{\theta}_k)$ does not play a role in the successive kinetostatic fold  compliance algorithm given by~\eqref{eq:KCMconv}.} Using this insight,  it is possible that one only  considers the sign of the torque vector $\pmb{\tau}(\pmb{\theta}_k)$ as an alternative to the heuristic approach in the conventional successive kinetostatic fold compliance in~\eqref{eq:KCMconv}. In particular, following the sign gradient descent optimization literature (see, e.g.,~\cite{pascanu2013difficulty,moulay2019properties}), we propose the \emph{SGD-based successive kinetostatic fold compliance algorithm} 
\begin{equation}
\pmb{\theta}_{k+1} = \pmb{\theta}_k + \kappa_k \, \text{sgn}\big(\pmb{\tau}(\pmb{\theta}_k)\big), \; k\in \mathbb{Z}_{0+}, 
\label{eq:sgd}
\end{equation}
where $\kappa_k$ is a step size that changes dynamically in every iteration and $\text{sgn}\big(\pmb{\tau}(\pmb{\theta}_k)\big)=\big[ \text{sgn}(\tau_1(\mathbf{x})),\cdots, \text{sgn}(\tau_{2N}(\mathbf{x})) \big]^\top$.  Furthermore, the step size $\kappa_k$ is tuned according to a proper adaptive step size strategy 
\begin{equation}
\kappa_{k+1} = \mathcal{S}(\kappa_k), \; k\in \mathbb{Z}_{0+},  
\label{eq:stepUpdate}
\end{equation}
where the mapping $\mathcal{S}: \mathbb{R}_+ \to \mathbb{R}_+$ should  satisfy $\mathcal{S}(\kappa_k) < \kappa_k$ for every positive $\kappa_k$. 

% Remarks on comparing the two
\begin{remark}
	Comparing the successive kinetostatic  fold compliance schemes given by~\eqref{eq:KCMconv} and~\eqref{eq:sgd}, it can be seen that none of them rely on the magnitude of the original torque vector $\pmb{\tau}(\pmb{\theta}_k)$.   Indeed, the conventional kinetostatic fold compliance in~\eqref{eq:KCMconv} relies on the normalized torque vector $\tfrac{\pmb{\tau}(\pmb{\theta}_k)}{| \pmb{\tau}(\pmb{\theta}_k) |_\infty}$, while the proposed SGD-based successive kinetostatic  fold compliance scheme in~\eqref{eq:sgd} relies on the sign of the torque vector, i.e., $\text{sgn}\big(\pmb{\tau}(\pmb{\theta}_k)\big)$.  Furthermore, while the step size $\kappa_0$ in~\eqref{eq:KCMconv} is fixed, the step size $\kappa_k$ in~\eqref{eq:sgd} varies in a dynamic manner. 
\end{remark}

To design the adaptive step size mapping $\mathcal{S}(\cdot)$ in~\eqref{eq:stepUpdate}, one can use various established methods such as the following step size adaptation rule (see, e.g.,~\cite{moulay2019properties})  
\begin{equation}
\mathcal{S}(\kappa_k) = \gamma_0 \kappa_k,\, k\in \mathbb{Z}_{0+},
\label{eq:dico}
\end{equation}
where $\gamma_0 \in (0,1)$ is a positive constant, resulting in the step size sequence $\big\{ \kappa_0\cdot (\gamma_0)^k \big\}_{k\in \mathbb{Z}_{0+}}$. It is remarked  that in the special case of $\gamma_0=0.5$, the step size adaptation rule in~\eqref{eq:dico} is called the DICHO algorithm.  

Moulay \emph{et al.}~\cite{moulay2019properties} have provided conditions on adaptive step size strategies under which sign gradient descent algorithms converge. Considering an unfolded conformation $\pmb{\theta}_0$ of a protein molecule in the vicinity of a folded conformation $\pmb{\theta}^\ast$, one can utilize Theorem~1 in~\cite{moulay2019properties} to find conditions on  the SGD-based successive fold compliance iteration in~\eqref{eq:sgd} with adaptive step size strategy given by~\eqref{eq:stepUpdate} to guarantee asymptotic convergence to the folded conformation $\pmb{\theta}^\ast$. In particular, assuming that $\pmb{\theta}^\ast$ is an isolated local minimum of $\mathcal{G}(\pmb{\theta})$ and that $(\pmb{\theta}^\ast - \pmb{\theta}_k)^\top \text{sgn}(\pmb{\tau}(\pmb{\theta}_k)) > 0$ in an open neighborhood of $\pmb{\theta}^\ast$, the conditions due to Moulay \emph{et al.}~\cite{moulay2019properties} in the context of SGD-based protein folding read as follows: 
\begin{enumerate}
	\item $0 < \kappa_k < 2(\pmb{\theta}^\ast - \pmb{\theta}_k)^\top \text{sgn}(\pmb{\tau}(\pmb{\theta}_k)) $;
	\item $\kappa_k (\pmb{\theta}^\ast - \pmb{\theta}_k)^\top \text{sgn}(\pmb{\tau}(\pmb{\theta}_k)) \geq c \, \| \pmb{\theta}^\ast - \pmb{\theta}_k\|^\alpha$ for some positive $\alpha$ and $c$; and, 
	\item $\lim\limits_{k \to \infty} \kappa_k = 0$.
\end{enumerate}

\section{Numerical Simulations}
\label{sec:sims}
In this section we present numerical simulation results for KCM-based protein folding dynamics to validate our proposed SGD-based successive kinetostatic fold compliance algorithm given by~\eqref{eq:sgd} and compare its performance against the conventional kinetostatic fold compliance algorithm given by~\eqref{eq:KCMconv}. 

In our simulations, we considered a protein molecule backbone chain consisting of $N-1=15$ peptide planes, which corresponds to having a $2N=32$-dimensional dihedral angle space (i.e., the conformation vector $\pmb{\theta}$ in~\eqref{eq:thetaConfig} consists of $32$ dihedral angles).  Our implementation followed the guidelines of Protofold~I~\cite{kazerounian2005protofold,Tavousi2016} on an Intel\textsuperscript{\textregistered} Core\textsuperscript{\texttrademark} i7-6770HQ CPU@2.60GHz. To demonstrate the advantages of our proposed SGD-based folding algorithm,  we purposefully chose a relatively large parameter $\kappa_0$ for the conventional kinetostatic fold compliance algorithm given by~\eqref{eq:KCMconv}. In particular, we set $\kappa_0=0.01$. Furthermore, we set the initial step size for the SGD-based algorithm in~\eqref{eq:sgd} to be equal to $\kappa_0=0.01$ (the same as the fixed step size in ~\eqref{eq:KCMconv}). Moreover, we chose an adaptive step size strategy similar to~\eqref{eq:dico} with $\gamma_0=0.99$. 

\begin{figure*}[t]
	\centering
	\includegraphics[width=0.7\textwidth]{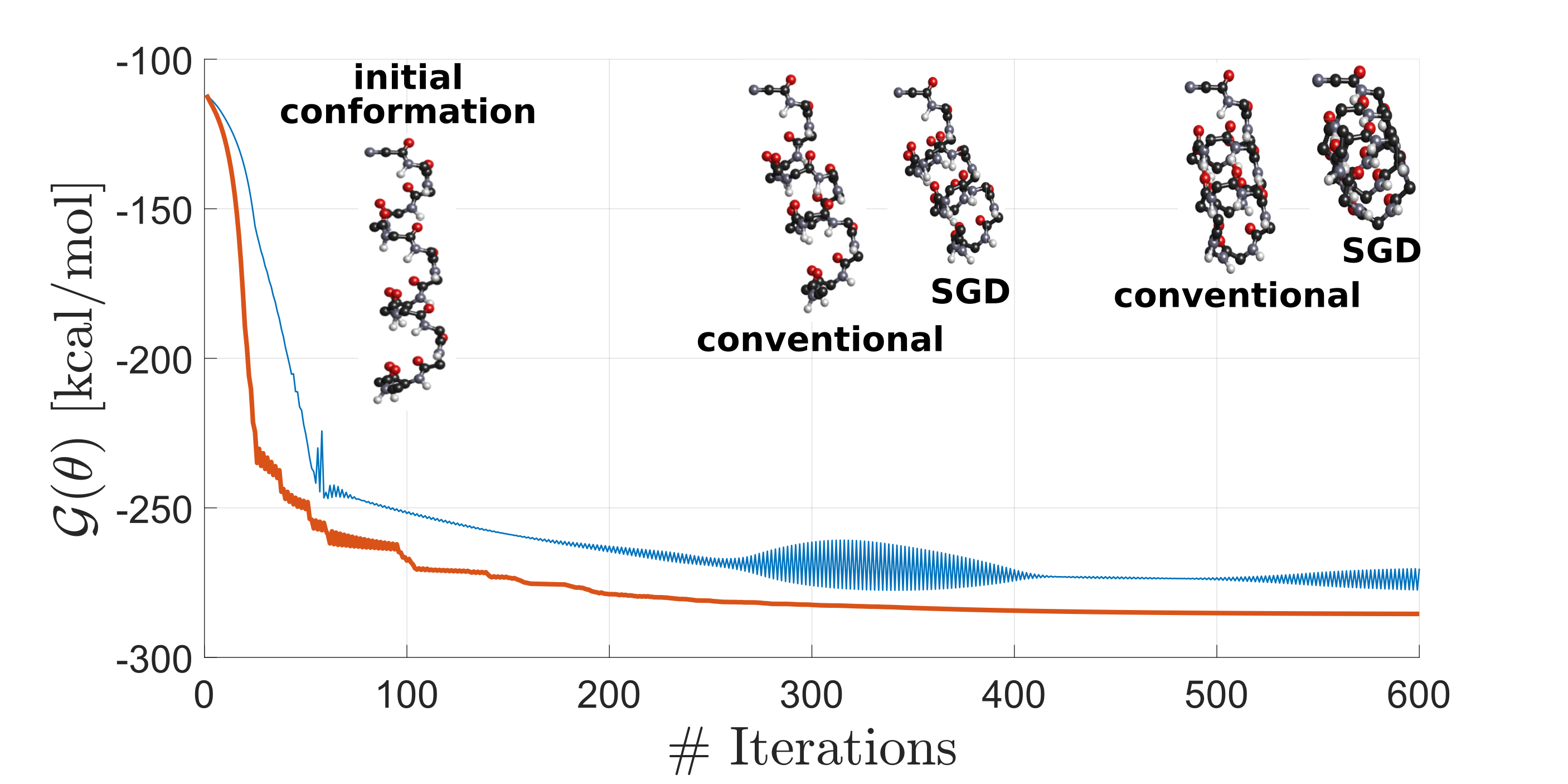}
	\caption{\small The free energy of the protein molecule under the conventional successive kinetostatic fold compliance algorithm (blue curve) and the proposed SGD-based algorithm (red curve).}
	%\vspace{-0.5ex}
	\label{fig:marss}
\end{figure*}

Our initial protein molecule conformations in both tests were chosen to be the same pre-coiled backbone chain in a vicinity of an $\alpha$-helix conformation. Figure~\ref{fig:marss} depicts the free energy of the protein backbone peptide chain starting from the same initial conformations (also depicted in the same figure) under the conventional algorithm (blue curve) and the SGD-based algorithm (red curve). Furthermore, Figure~\ref{fig:marss} depicts the configurations of the protein molecule in the  30\textsuperscript{th} and the 600\textsuperscript{th} iterations under the conventional and SGD-based folding algorithms, respectively.

As it can be clearly seen from Figure~\ref{fig:marss} the free energy under the successive kinetostatic fold compliance algorithm has gone through oscillations and converged to a higher free energy level of the protein molecule. On the other hand, the free energy of the protein backbone chain under the SGD-based algorithm has not gone through the same oscillations and converged to a lower free energy level of the protein molecule in a faster manner. 

\begin{figure}[t]
	\centering
	\includegraphics[width=0.48\textwidth]{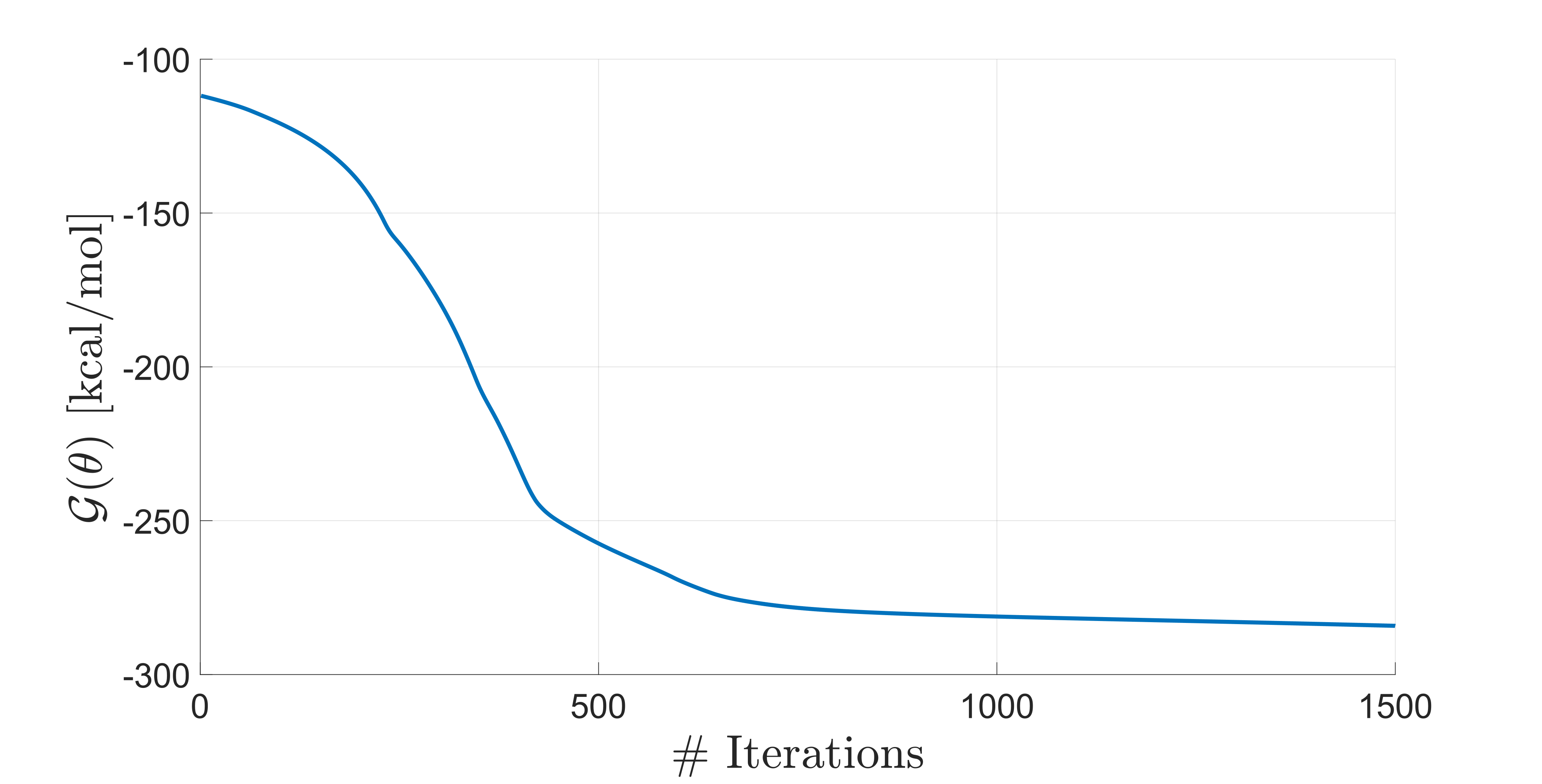}
	\caption{\small The free energy of the protein molecule under the conventional successive kinetostatic fold compliance algorithm with a smaller step size ($\kappa_0=0.001$) manages to converge to the same free energy level after a higher number of iterations (1500 iterations in contrast with 600 iterations).}
	%\vspace{-0.5ex}
	\label{fig:marss2a}
\end{figure}

The observations in Figure~\ref{fig:marss} are in accordance with a well known fact from the established KCM literature that the price to pay for convergence is to choose smaller step sizes with more iterations required for convergence and a consequent higher computational burden. To demonstrate this fact, we reduced the step size associated with the conventional algorithm from $\kappa_0=0.01$ to $\kappa_0=0.001$. The free energy level of the protein backbone chain is depicted in Figure~\ref{fig:marss2a}. As it can be seen from the figure, the free energy of the protein molecule under the conventional successive kinetostatic fold compliance algorithm with a smaller step size ($\kappa_0=0.001$) manages to converge to the same free energy level as of its SGD-based counterpart, but only with a higher number of numerical iterations (1500 iterations in contrast with 600 iterations). 

\section{Concluding Remarks and Future Research Directions}
\label{sec:conclusion}

In a departure from the established kinetostatic fold compliance literature on numerically simulating  the protein folding process, this paper proposed a sign gradient descent algorithm for predicting the three-dimensional folded protein molecule structures. The more stable and robust convergence properties of the proposed SGD-based algorithm makes it suitable for accurate simulation of the range of motion of peptide-based nanorobots/nanomachines  such as parallel nano-mechanisms~\cite{hamdi2005molecular,hamdi2009multiscale} and  closed-loop cyclic 7-R peptide-based mechanisms~\cite{mundrane2022exploring}. As future research directions, we envision that the proposed SGD-based successive kinetostatic fold compliance literature can be utilized for efficient numerical investigation of the KCM-based protein folding dynamics under solvation effects and  entropy-loss constraints. Furthermore, our proposed algorithm has the potential of lending itself to stochastic SGD extensions by relying on the emerging literature of stochastic sign descent methods (see, e.g.,~\cite{safaryan2021stochastic}). 

\section*{Acknowledgments}
This work is supported by the National Science Foundation (NSF) through the award number CMMI-2153744. 

\bibliographystyle{IEEEtran}
\bibliography{./elsBibIntegrator.bib}

\end{document}